# Split quaternions and time-like constant slope surfaces in Minkowski 3-space

Murat Babaarslan[a,*] Yusuf Yayli[b]

[a]Department of Mathematics, Bozok University, 66100, Yozgat-Turkey
[b]Department of Mathematics, Ankara University, 06100, Ankara-Turkey

**Abstract**

In this paper, we prove that time-like constant slope surfaces can be reparametrized by using rotation matrices corresponding to unit time-like split quaternions and homothetic motions. Afterwards we give some examples to illustrate our main results by using Mathematica.

## 1. Introduction

We can find a lot of methods to construct rotations, for example; orthogonal matrices, Euler angles, Rodrigues rotational formulae which is a good solution to rotation about an arbitrary axis, see [7] and quaternions. In between them, quaternions are more efficient tools for describing rotations about an arbitrary axis since we can obtain easily rotation matrices by using quaternions. Thus they are usually used in robotics, computer graphics, animations, aerospace applications, fractals and virtual reality (see [1], [11]).

In Computer Aided Geometric Design (CAGD), B-Spline and NURBS curves are well known geometric modeling tools. Also trigonometric curves can construct a lot of important curves due to their trigonometric basis functions, for example; circle, circular cylinder, rotational surfaces and helix which makes a constant angle with a fixed direction, see [13] and references therein. If we want to give a different example, certainly, we should think constant slope surface which makes a constant angle with the position vector. These surfaces can be thought as a generalization of helices which arise in DNA double, collagen triple helix, nano-springs, carbon nano-tubes, helical staircases, helical structures in fractal geometry and so on (see [5]).

Munteanu [9] showed that a constant slope surface can be constructed by using a unit speed curve on the Euclidean 2-sphere $\mathbb{S}^2$ and an equiangular spiral.

We know that a rotational surface can be obtained by using a rotation matrix. In [1], we demonstrated that constant slope surfaces can be obtained by using quaternions. After that in [2], we found the relations between space-like constant slope surfaces [3] and split quaternions in Minkowski 3-space $\mathbb{R}^3_1$.

Fu and Wang [4] studied time-like constant slope surfaces and classified these surfaces in $\mathbb{R}^3_1$. They showed that $S \subset \mathbb{R}^3_1$ is a time-like constant slope surface lying in the time-like cone if and only if it can be parametrized by

$$x(u,v) = u\cosh\theta \left(\cosh\xi_1(u)f(v) + \sinh\xi_1(u)f(v) \wedge f'(v)\right), \tag{1.1}$$

where, $\theta$ is a positive constant angle function, $\xi_1(u) = \tanh\theta \ln u$ and $f$ is a unit speed space-like curve on pseudo-hyperbolic space $\mathbb{H}^2$.

$S \subset \mathbb{R}^3_1$ is a time-like constant slope surface lying in the space-like cone if and only if it can be parametrized by

$$x(u,v) = u\sin\theta \left(\cos\xi_2(u)g(v) + \sin\xi_2(u)g(v) \wedge g'(v)\right), \tag{1.2}$$

where, $\theta$ is a positive constant angle function satisfying $\theta = (0, \pi/2]$, $\xi_2(u) = \cot\theta \ln u$ and $g$ is a unit speed time-like curve on pseudo-sphere $\mathbb{S}^2_1$.

Also, $S \subset \mathbb{R}^3_1$ is a time-like constant slope surface lying in the space-like cone if and only if it can be parametrized by

$$x(u,v) = u \sinh\theta \left(\cosh\xi_3(u)h(v) + \sinh\xi_3(u)h(v) \wedge h'(v)\right), \tag{1.3}$$

where, $\theta$ is a positive constant angle function, $\xi_3(u) = \coth\theta \ln u$ and $h$ is a unit speed space-like curve on pseudo-sphere $\mathbb{S}_1^2$.

In light of recent studies, the purpose of this paper is to give some relations between split quaternions and time-like constant slope surfaces in $\mathbb{R}_1^3$. We will show that time-like constant slope surfaces can be reparametrized by using rotation matrices corresponding to unit time-like split quaternions and homothetic motions. Also we will give some examples to strengthen the both this paper and [4] by using Mathematica.

## 2. Preliminaries

In this section, we recall some important concepts and formulas regarding split quaternions and semi-Euclidean spaces.

A split quaternion $p$ can be written as

$$p = p_1 \mathbf{1} + p_2 \mathbf{i} + p_3 \mathbf{j} + p_4 \mathbf{k},$$

where $p_1, p_2, p_3, p_4 \in \mathbb{R}$ and $\mathbf{i}, \mathbf{j}, \mathbf{k}$ are split quaternion units which satisfy the non-commutative multiplication rules

$$\mathbf{i}^2 = -1, \ \mathbf{j}^2 = \mathbf{k}^2 = 1, \ \mathbf{i} \times \mathbf{j} = -\mathbf{j} \times \mathbf{i} = \mathbf{k}, \ \mathbf{j} \times \mathbf{k} = -\mathbf{k} \times \mathbf{j} = -\mathbf{i} \text{ and } \mathbf{k} \times \mathbf{i} = -\mathbf{i} \times \mathbf{k} = \mathbf{j}.$$

Let us denote the algebra of split quaternions by $\mathbb{H}'$ and its natural basis by $\{1, \mathbf{i}, \mathbf{j}, \mathbf{k}\}$. An element of $\mathbb{H}'$ is called a split quaternion.

For a split quaternion $p = p_1 \mathbf{1} + p_2 \mathbf{i} + p_3 \mathbf{j} + p_4 \mathbf{k}$, the conjugate $\bar{p}$ of $p$ is defined by

$$\bar{p} = p_1 \mathbf{1} - p_2 \mathbf{i} - p_3 \mathbf{j} - p_4 \mathbf{k}$$

[6]. Scalar and vector parts of a split quaternion $p$ are denoted by $S_p = p_1$ and $V_p = p_2 \mathbf{i} + p_3 \mathbf{j} + p_4 \mathbf{k}$, respectively. The split quaternion product of two split quaternions $p = (p_1, p_2, p_3, p_4)$ and $q = (q_1, q_2, q_3, q_4)$ is defined as

$$p \times q = p_1 q_1 + <V_p, V_q> + p_1 V_q + q_1 V_p + V_p \wedge V_q, \tag{2.1}$$

where

$$<V_p, V_q> = -p_2q_2 + p_3q_3 + p_4q_4$$

and

$$V_p \wedge V_q = \begin{vmatrix} -\mathbf{i} & \mathbf{j} & \mathbf{k} \\ p_2 & p_3 & p_4 \\ q_2 & q_3 & q_4 \end{vmatrix} = (p_4q_3 - p_3q_4)\mathbf{i} + (p_4q_2 - p_2q_4)\mathbf{j} + (p_2q_3 - p_3q_2)\mathbf{k}$$

If $S_p = 0$, then $p$ is called as a pure split quaternion [11].

**Definition 2.1.** If $\mathbb{R}^n$ is equipped with the metric tensor

$$<\mathbf{u}, \mathbf{v}> = -\sum_{i=1}^{v} u_i v_i + \sum_{i=v+1}^{n} u_i v_i,$$

it is called semi-Euclidean space and denoted by $\mathbb{R}^n_v$, where $v$ is called the index of the metric, $0 \leq v \leq n$ and $\mathbf{u}, \mathbf{v} \in \mathbb{R}^n$. If $v = 0$, semi-Euclidean space $\mathbb{R}^n_v$ is reduced to $\mathbb{R}^n$. For $n \geq 2$, $\mathbb{R}^n_1$ is called Minkowski $n$-space [10].

**Definition 2.2.** A vector $\mathbf{w} \in \mathbb{R}^n_v$ is called
(i) space-like if $<\mathbf{w}, \mathbf{w}> > 0$ or $\mathbf{w} = 0$,
(ii) time-like if $<\mathbf{w}, \mathbf{w}> < 0$,
(iii) light-like (null) if $<\mathbf{w}, \mathbf{w}> = 0$ and $\mathbf{w} \neq 0$.

The norm of a vector $\mathbf{w} \in \mathbb{R}^n_v$ is $\sqrt{|<\mathbf{w}, \mathbf{w}>|}$. Two vectors $\mathbf{w}_1$ and $\mathbf{w}_2$ in $\mathbb{R}^n_v$ are said to be orthogonal if $<\mathbf{w}_1, \mathbf{w}_2> = 0$ [10].

We can define pseudo-sphere and pseudo-hyperbolic space in $\mathbb{R}^n_v$ as follows:

$$\mathbb{S}^{n-1}_v = \left\{ (v_1, \ldots, v_n) \in \mathbb{R}^n_v \middle| -\sum_{i=1}^{v} v_i^2 + \sum_{i=v+1}^{n} v_i^2 = 1 \right\},$$

and

$$\mathbb{H}^{n-1}_{v-1} = \left\{ (v_1, \ldots, v_n) \in \mathbb{R}^n_v \middle| -\sum_{i=1}^{v} v_i^2 + \sum_{i=v+1}^{n} v_i^2 = -1 \right\}.$$

Also, for $v = 1$ and $v_1 > 0$, $\mathbb{H}^{n-1} = \mathbb{H}^{n-1}_0$ is called a pseudo-hyperbolic space of $\mathbb{R}^n_1$.

It can be shown that $-p \times \bar{p} = -p_1^2 - p_2^2 + p_3^2 + p_4^2$. Therefore we identify $\mathbb{H}'$ with semi-Euclidean space $\mathbb{R}_2^4$ [6].

Thus we can define time-like, space-like and light-like split quaternions as follows:

**Definition 2.3.** A split quaternion $p$ is space-like, time-like or light-like, if $I_p < 0$, $I_p > 0$ or $I_p = 0$, respectively, where $I_p = p_1^2 + p_2^2 - p_3^2 - p_4^2$ [11].

**Definition 2.4.** The norm of $p = (p_1, p_2, p_3, p_4)$ is defined as
$$N_p = \sqrt{\left|p_1^2 + p_2^2 - p_3^2 - p_4^2\right|}.$$
If $N_p = 1$, then $p$ is called as a unit split quaternion and $p_0 = p/N_p$ is a unit split quaternion for $N_p \neq 0$. Also space-like and time-like split quaternions have multiplicative inverses having the property $p \times p^{-1} = p^{-1} \times p = 1$ and they are constructed by $p^{-1} = \bar{p}/I_p$. Light-like split quaternions have no inverses [11].

The set of space-like split quaternions is not a group because it is not closed under multiplication. Therefore we are only interested in time-like split quaternions. The vector part of any time-like split quaternion can be space-like or time-like. Trigonometric forms of them are as below:

(i) Every time-like split quaternion with the space-like vector part can be written as
$$p = N_p(\cosh\theta + w\sinh\theta),$$
where $w$ is a unit space-like vector in $\mathbb{R}_1^3$.

(ii) Every time-like split quaternion with the time-like vector part can be written as
$$p = N_p(\cos\theta + w\sin\theta),$$
where $w$ is a unit time-like vector in $\mathbb{R}_1^3$ [8, 11].

Unit time-like split quaternions are used to construct rotations in $\mathbb{R}_1^3$:

If $p = (p_1, p_2, p_3, p_4)$ is a unit time-like split quaternion, using the transformation law $(p \times V \times p^{-1})_i = \sum_{j=1}^{3} R_{ij} V_j$, the corresponding rotation matrix can be found as

$$R_p = \begin{bmatrix} p_1^2 + p_2^2 + p_3^2 + p_4^2 & 2p_1p_4 - 2p_2p_3 & -2p_1p_3 - 2p_2p_4 \\ 2p_2p_3 + 2p_4p_1 & p_1^2 - p_2^2 - p_3^2 + p_4^2 & -2p_3p_4 - 2p_2p_1 \\ 2p_2p_4 - 2p_3p_1 & 2p_2p_1 - 2p_3p_4 & p_1^2 - p_2^2 + p_3^2 - p_4^2 \end{bmatrix}, \quad (2.2)$$

where $V$ is a split quaternion. This is possible with unit time-like split quaternions. Also causal character of vector part of the unit time-like split quaternion $p$ is important. If the vector part of $p$ is time-like or space-like then the rotation angle is spherical or hyperbolic, respectively [11].

In $\mathbb{R}_1^3$, one-parameter homothetic motion of a body is generated by the transformation

$$\begin{bmatrix} Y \\ 1 \end{bmatrix} = \begin{bmatrix} hA & C \\ 0 & 1 \end{bmatrix} \begin{bmatrix} X \\ 1 \end{bmatrix},$$

where $X$ and $C$ are real matrices of $3 \times 1$ type and $h$ is a homothetic scale and $A \in SO_1(3)$. $A$, $h$ and $C$ are differentiable functions of $C^\infty$ class of a parameter $t$ (see [12]).

## 3. Split quaternions and time-like constant slope surfaces lying in the time-like cone

A unit time-like split quaternion with the space-like vector part

$$Q(u,v) = \cosh(\xi_1(u)/2) - \sinh(\xi_1(u)/2) f'(v)$$

defines a 2-dimensional surface on pseudo-hyperbolic space $\mathbb{H}_1^3$, where $\xi_1(u) = \tanh\theta \ln u$, $\theta$ is a positive constant angle function, $f' = (f_1', f_2', f_3')$ and $f$ is a unit speed space-like curve on $\mathbb{H}^2$. Thus, using (2.2), the corresponding rotation matrix can be found as

$$R_Q = \begin{bmatrix} \cosh^2\frac{\xi_1}{2} + \sinh^2\frac{\xi_1}{2}(f_1'^2 + f_2'^2 + f_3'^2) & -2\sinh^2\frac{\xi_1}{2}f_1'f_2' - \sinh\xi_1 f_3' & -2\sinh^2\frac{\xi_1}{2}f_1'f_3' + \sinh\xi_1 f_2' \\ 2\sinh^2\frac{\xi_1}{2}f_1'f_2' - \sinh\xi_1 f_3' & \cosh^2\frac{\xi_1}{2} + \sinh^2\frac{\xi_1}{2}(-f_1'^2 - f_2'^2 + f_3'^2) & -2\sinh^2\frac{\xi_1}{2}f_2'f_3' + \sinh\xi_1 f_1' \\ 2\sinh^2\frac{\xi_1}{2}f_1'f_3' + \sinh\xi_1 f_2' & -2\sinh^2\frac{\xi_1}{2}f_2'f_3' - \sinh\xi_1 f_1' & \cosh^2\frac{\xi_1}{2} + \sinh^2\frac{\xi_1}{2}(-f_1'^2 + f_2'^2 - f_3'^2) \end{bmatrix}.$$

(3.1)

Now we give the relation between unit time-like split quaternions with the space-like vector parts and time-like constant slope surfaces lying in the time-like cone.

**Theorem 3.1.** Let $x: S \to \mathbb{R}_1^3$ be a time-like constant slope surface immersed in Minkowski 3-space $\mathbb{R}_1^3$ and $x$ lies in the time-like cone. Then the time-like constant slope surface $S$ can be reparametrized by

$$x(u,v) = Q_1(u,v) \times Q_2(u,v),$$

where $Q_1(u,v) = \cosh \xi_1(u) - \sinh \xi_1(u) f'(v)$ is a unit time-like split quaternion with the space-like vector part, $Q_2(u,v) = u \cosh \theta f(v)$ is a surface and a pure split quaternion in $\mathbb{R}_1^3$ and $f$ is a unit speed space-like curve on $\mathbb{H}^2$.

**Proof.** Since $Q_1(u,v) = \cosh \xi_1(u) - \sinh \xi_1(u) f'(v)$ and $Q_2(u,v) = u \cosh \theta f(v)$, we have

$$\begin{aligned} Q_1(u,v) \times Q_2(u,v) &= \left( \cosh \xi_1(u) - \sinh \xi_1(u) f'(v) \right) \times \left( u \cosh \theta f(v) \right) \\ &= u \cosh \theta \left( \cosh \xi_1(u) - \sinh \xi_1(u) f'(v) \right) \times f(v) \\ &= u \cosh \theta \cosh \xi_1(u) f(v) - u \cosh \theta \sinh \xi_1(u) f'(v) \times f(v). \end{aligned} \quad (3.2)$$

By using (2.1), we have

$$f' \times f = <f', f> + f' \wedge f.$$

Since $f$ is a unit speed space-like curve on $\mathbb{H}^2$, we get

$$-f' \times f = f \wedge f'. \quad (3.3)$$

Substituting (3.3) into (3.2), we obtain

$$Q_1(u,v) \times Q_2(u,v) = u \cosh \theta \left( \cosh \xi_1(u) f(v) + \sinh \xi_1(u) f(v) \wedge f'(v) \right).$$

Thus, using (1.1), we conclude that

$$Q_1(u,v) \times Q_2(u,v) = x(u,v).$$

This completes the proof.

From Theorem 3.1, we can see that the time-like constant slope surface $S$ lying in the time-like cone is the split quaternion product of 2-dimensional surfaces $Q_1(u,v)$ on $\mathbb{H}_1^3$ and $Q_2(u,v)$ in $\mathbb{R}_1^3$.

We now consider the relations among rotation matrices $R_Q$, homothetic motions and time-like constant slope surfaces lying in the time-like cone. We have the following results of Theorem 3.1.

**Corollary 3.2.** Let $R_Q$ be the rotation matrix corresponding to the unit time-like split quaternion with the space-like vector part $Q(u,v)$. Then the time-like constant slope surface $S$ can be written as

$$x(u,v) = R_Q Q_2(u,v).$$

**Corollary 3.3.** For the homothetic motion $\tilde{Q}(u,v) = u\cosh\theta Q_1(u,v)$, the time-like constant slope surface $S$ can be reparametrized by $x(u,v) = \tilde{Q}(u,v) \times f(v)$. Therefore, we have

$$x(u,v) = u\cosh\theta R_Q f(v). \tag{3.4}$$

Thus, we can give the following example:

**Example 3.4.** We consider a unit space-like curve on $\mathbb{H}^2$ defined by

$$f(v) = (\cosh v, 0, \sinh v).$$

If we take $\theta = 7$ and $\tanh 7 \cong 1$, the homothetic motion is

$$\tilde{Q}(u,v) = u\cosh 7 \big(\cosh(\ln u) + (-\sinh(\ln u)\sinh v, 0, -\sinh(\ln u)\cosh v)\big).$$

Using (3.4), we have the following time-like constant slope surface lying in the time-like cone:

$$x(u,v) = u\cosh 7 \begin{bmatrix} \cosh^2\left(\frac{\ln u}{2}\right) + \sinh^2\left(\frac{\ln u}{2}\right)\cosh 2v & -\sinh(\ln u)\cosh v & -\sinh^2\left(\frac{\ln u}{2}\right)\sinh 2v \\ -\sinh(\ln u)\cosh v & \cosh(\ln u) & \sinh(\ln u)\sinh v \\ \sinh^2\left(\frac{\ln u}{2}\right)\sinh 2v & -\sinh(\ln u)\sinh v & \cosh^2\left(\frac{\ln u}{2}\right) - \sinh^2\left(\frac{\ln u}{2}\right)\cosh 2v \end{bmatrix} \begin{bmatrix} \cosh v \\ 0 \\ \sinh v \end{bmatrix}$$

and then

$$x(u,v) = \begin{bmatrix} u\cosh 7\cosh(\ln u)\cosh v \\ -u\cosh 7\sinh(\ln u) \\ u\cosh 7\cosh(\ln u)\sinh v \end{bmatrix}.$$

Thus, the picture of $x(u,v) = u\cosh\theta R_Q f(v)$:

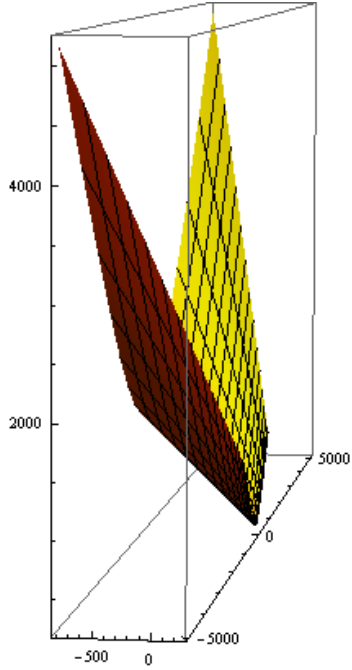

Figure 1: Time-like constant slope surface lying in the time-like cone $x(u,v) = u\cosh\theta R_Q f(v)$, $f(v) = (\cosh v, 0, \sinh v)$, $\theta = 7$.

We can give the following remarks regarding Theorem 3.1 and Corollary 3.3.

**Remark 3.5.** Theorem 3.1 says that the position vectors on the surface $Q_2(u,v)$ are rotated by $Q_1(u,v)$ through the hyperbolic angle $\xi_1(u)$ about the space-like axis $\text{span}\{f'(v)\}$.

**Remark 3.6.** Corollary 3.3 says that the position vector of the space-like curve $f(v)$ is rotated by $\tilde{Q}(u,v)$ through the hyperbolic angle $\xi_1(u)$ about the space-like axis $\text{span}\{f'(v)\}$ and extended through the homothetic scale $u\cosh\theta$.

## 4. Split quaternions and time-like constant slope surfaces lying in the space-like cone

As Section 3, we consider a unit time-like split quaternion with the time-like vector part

$$Q(u,v) = \cos(\xi_2(u)/2) - \sin(\xi_2(u)/2)g'(v),$$

where $\xi_2(u) = \cot\theta \ln u$, $\theta$ is a positive constant angle function satisfying $\theta = (0, \pi/2]$, $g' = (g_1', g_2', g_3')$ and $g$ is a unit speed time-like curve on $\mathbb{S}_1^2$. Thus, using (2.2), the corresponding rotation matrix can be found as

$$R_Q = \begin{bmatrix} \cos^2\frac{\xi_2}{2} + \sin^2\frac{\xi_2}{2}(g_1'^2 + g_2'^2 + g_3'^2) & -2\sin^2\frac{\xi_2}{2}g_1'g_2' - \sin\xi_2 g_3' & -2\sin^2\frac{\xi_2}{2}g_1'g_3' + \sin\xi_2 g_2' \\ 2\sin^2\frac{\xi_2}{2}g_1'g_2' - \sin\xi_2 g_3' & \cos^2\frac{\xi_2}{2} + \sin^2\frac{\xi_2}{2}(-g_1'^2 - g_2'^2 + g_3'^2) & -2\sin^2\frac{\xi_2}{2}g_2'g_3' + \sin\xi_2 g_1' \\ 2\sin^2\frac{\xi_2}{2}g_1'g_3' + \sin\xi_2 g_2' & -2\sin^2\frac{\xi_2}{2}g_2'g_3' - \sin\xi_2 g_1' & \cos^2\frac{\xi_2}{2} + \sin^2\frac{\xi_2}{2}(-g_1'^2 + g_2'^2 - g_3'^2) \end{bmatrix},$$

(4.1)

Now we give the relation between unit time-like quaternions with the time-like vector parts and the time-like constant slope surfaces lying in the space-like cone.

**Theorem 4.1.** Let $x: S \to \mathbb{R}_1^3$ be a time-like constant slope surface immersed in Minkowski 3-space $\mathbb{R}_1^3$ and $x$ lies in the space-like cone. Then the time-like constant slope surface $S$ can be reparametrized by

$$x(u,v) = Q_1(u,v) \times Q_2(u,v),$$

where $Q_1(u,v) = \cos\xi_2(u) - \sin\xi_2(u)g'(v)$ is a unit time-like split quaternion with time-like vector part, $Q_2(u,v) = u\sin\theta g(v)$ is a surface and a pure split quaternion in $\mathbb{R}_1^3$ and $g$ is a unit speed time-like curve on $\mathbb{S}_1^2$.

**Proof.** Since $Q_1(u,v) = \cos\xi_2(u) - \sin\xi_2(u)g'(v)$ and $Q_2(u,v) = u\sin\theta g(v)$, we obtain

$$\begin{aligned} Q_1(u,v) \times Q_2(u,v) &= \left(\cos\xi_2(u) - \sin\xi_2(u)g'(v)\right) \times \left(u\sin\theta g(v)\right) \\ &= u\sin\theta\left(\cos\xi_2(u) - \sin\xi_2(u)g'(v)\right) \times g(v) \\ &= u\sin\theta\cos\xi_2(u)g(v) - u\sin\theta\sin\xi_2(u)g'(v) \times g(v). \end{aligned}$$

(4.2)

By using (2.1), we have

$$g' \times g = <g', g> + g' \wedge g.$$

Since $g$ is a unit speed time-like curve on $\mathbb{S}_1^2$, we have

$$-g' \times g = g \wedge g'.$$

(4.3)

Substituting (4.3) into (4.2) gives

$$Q_1(u,v) \times Q_2(u,v) = u\sin\theta\left(\cos\xi_2(u)g(v) + \sin\xi_2(u)g(v)\wedge g'(v)\right).$$

Therefore applying (1.2), we conclude that

$$Q_1(u,v) \times Q_2(u,v) = x(u,v).$$

Thus the proof is completed.

From Theorem 4.1, we can see that the time-like constant slope surface $S$ lying in the space-like cone is the split quaternion product of 2-dimensional surfaces $Q_1(u,v)$ on $\mathbb{H}_1^3$ and $Q_2(u,v)$ in $\mathbb{R}_1^3$.

Let us consider the relations among rotation matrices $R_Q$, homothetic motions and time-like constant slope surfaces lying in the space-like cone. We have the following results of Theorem 4.1.

**Corollary 4.2.** Let $R_Q$ be the rotation matrix corresponding to the unit time-like split quaternion with time-like vector part $Q(u,v)$. Then the time-like constant slope surface $S$ can be written as

$$x(u,v) = R_Q Q_2(u,v).$$

**Corollary 4.3.** For the homothetic motion $\tilde{Q}(u,v) = u\sin\theta Q_1(u,v)$, the time-like constant slope surface $S$ can be reparametrized by $x(u,v) = \tilde{Q}(u,v) \times g(v)$. Therefore, we have

$$x(u,v) = u\sin\theta R_Q g(v). \tag{4.4}$$

Also, we can give an example:

**Example 4.4.** We consider a unit time-like curve on $\mathbb{S}_1^2$ defined by

$$g(v) = (\sinh v, 0, \cosh v).$$

If we take $\theta = \pi/4$, the homothetic motion is

$$\tilde{Q}(u,v) = \frac{\sqrt{2}}{2}u\left(\cos(\ln u) + (-\sin(\ln u)\cosh v, 0, -\sin(\ln u)\sinh v)\right).$$

By using (4.4), we have the following time-like constant slope surface lying in the space-like cone:

$$x(u,v)=\frac{\sqrt{2}}{2}u\begin{bmatrix}\cos^2\left(\frac{\ln u}{2}\right)+\sin^2\left(\frac{\ln u}{2}\right)\cosh 2v & \sin\left(\frac{\ln u}{2}\right)\sinh v & -\sin^2\left(\frac{\ln u}{2}\right)\sinh 2v \\ \sin(\ln u)\sinh v & \cosh(\ln u) & -\sin(\ln u)\cosh v \\ \sin^2\left(\frac{\ln u}{2}\right)\sinh 2v & \sin(\ln u)\cosh v & \cos^2\left(\frac{\ln u}{2}\right)-\sin^2\left(\frac{\ln u}{2}\right)\cosh 2v\end{bmatrix}\begin{bmatrix}\sinh v \\ 0 \\ \cosh v\end{bmatrix}$$

and then

$$x(u,v)=\begin{bmatrix}\frac{\sqrt{2}}{2}u\cos(\ln u)\sinh v \\ \frac{\sqrt{2}}{2}u\sin(\ln u) \\ \frac{\sqrt{2}}{2}u\cos(\ln u)\cosh v\end{bmatrix}.$$

Thus, the picture of $x(u,v)=u\sin\theta R_Q g(v)$:

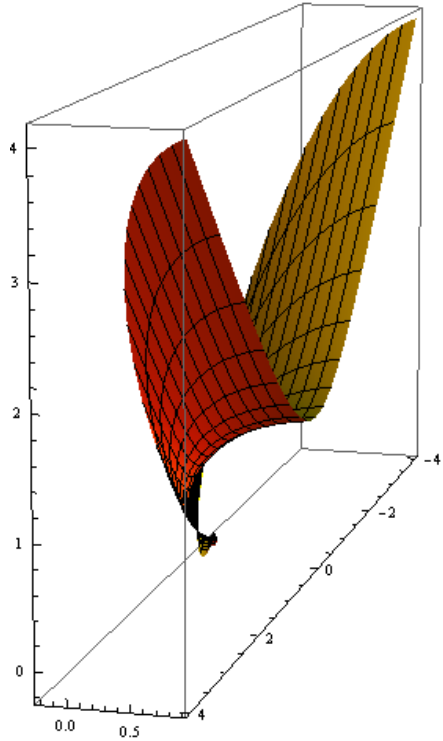

Figure 2: Time-like constant slope surface lying in the space-like cone $x(u,v)=u\sin\theta R_Q g(v)$, $g(v)=(\sinh v, 0, \cosh v)$, $\theta=\pi/4$.

We can give the following remarks regarding Theorem 4.1 and Corollary 4.3.

**Remark 4.5.** Theorem 4.1 says that the position vectors on the surface $Q_2(u,v)$ are rotated by $Q_1(u,v)$ through the spherical angle $\xi_2(u)$ about the time-like axis $\text{span}\{g'(v)\}$.

**Remark 4.6.** Corollary 4.3 says that the position vector of the time-like curve $g(v)$ is rotated by $\tilde{Q}(u,v)$ through the spherical angle $\xi_2(u)$ about the time-like axis $\text{span}\{g'(v)\}$ and extended through the homothetic scale $u\sin\theta$.

Also, we can give the relation between unit time-like split quaternions with the space-like vector parts and the time-like constant slope surfaces lying in the space-like cone. We consider a unit time-like split quaternion with the space-like vector part

$$Q(u,v) = \cosh(\xi_3(u)/2) - \sinh(\xi_3(u)/2)h'(v),$$

where $\xi_3(u) = \coth\theta \ln u$, $\theta$ is a positive constant angle function, $h' = (h'_1, h'_2, h'_3)$ and $h$ is a unit speed space-like curve on $\mathbb{S}_1^2$. Thus, using (2.2), the corresponding rotation matrix can be found as

$$R_Q = \begin{bmatrix} \cosh^2\frac{\xi_3}{2} + \sinh^2\frac{\xi_3}{2}(h_1'^2 + h_2'^2 + h_3'^3) & -2\sinh^2\frac{\xi_3}{2}h_1'h_2' - \sinh\xi_3 h_3' & -2\sinh^2\frac{\xi_3}{2}h_1'h_3' + \sinh\xi_3 h_2' \\ 2\sinh^2\frac{\xi_3}{2}h_1'h_2' - \sinh\xi_3 h_3' & \cosh^2\frac{\xi_3}{2} + \sinh^2\frac{\xi_3}{2}(-h_1'^2 - h_2'^2 + h_3'^3) & -2\sinh^2\frac{\xi_3}{2}h_2'h_3' + \sinh\xi_3 h_1' \\ 2\sinh^2\frac{\xi_3}{2}h_1'h_3' + \sinh\xi_3 h_2' & -2\sinh^2\frac{\xi_3}{2}h_2'h_3' - \sinh\xi_3 h_1' & \cosh^2\frac{\xi_3}{2} + \sinh^2\frac{\xi_3}{2}(-h_1'^2 + h_2'^2 - h_3'^3) \end{bmatrix},$$

(4.5)

**Theorem 4.7.** Let $x: S \to \mathbb{R}_1^3$ be a time-like constant slope surface immersed in Minkowski 3-space $\mathbb{R}_1^3$ and $x$ lies in the space-like cone. Then the time-like constant slope surface $S$ can be reparametrized by

$$x(u,v) = Q_1(u,v) \times Q_2(u,v),$$

where $Q_1(u,v) = \cosh\xi_3(u) - \sinh\xi_3(u)h'(v)$ is a unit time-like split quaternion with space-like vector part, $Q_2(u,v) = u\sinh\theta h(v)$ is a surface and a pure split quaternion in $\mathbb{R}_1^3$ and $h$ is a unit speed space-like curve on $\mathbb{S}_1^2$.

**Proof.** Since $Q_1(u,v) = \cosh \xi_3(u) - \sinh \xi_3(u) h'(v)$ and $Q_2(u,v) = u \sinh \theta h(v)$, we obtain

$$\begin{aligned} Q_1(u,v) \times Q_2(u,v) &= \left( \cosh \xi_3(u) - \sinh \xi_3(u) h'(v) \right) \times \left( u \sinh \theta h(v) \right) \\ &= u \sinh \theta \left( \cosh \xi_3(u) - \sinh \xi_3(u) h'(v) \right) \times h(v) \\ &= u \sinh \theta \cosh \xi_3(u) h(v) - u \sinh \theta \sinh \xi_3(u) h'(v) \times h(v). \end{aligned} \quad (4.6)$$

By using (2.1), we get

$$h' \times h = <h',h> + h' \wedge h.$$

Since $h$ is a unit speed space-like curve on $\mathbb{S}_1^2$, we have

$$-h' \times h = h \wedge h'. \quad (4.7)$$

Substituting (4.7) into (4.6) gives

$$Q_1(u,v) \times Q_2(u,v) = u \sinh \theta \left( \cosh \xi_3(u) h(v) + \sinh \xi_3(u) h(v) \wedge h'(v) \right).$$

Therefore applying (1.3) we conclude that

$$Q_1(u,v) \times Q_2(u,v) = x(u,v).$$

This completes the proof.

We have the following results of Theorem 4.7.

**Corollary 4.8.** Let $R_Q$ be the rotation matrix corresponding to the unit time-like split quaternion with space-like vector part $Q(u,v)$. Then the time-like constant slope surface $S$ can be written as

$$x(u,v) = R_Q Q_2(u,v).$$

**Corollary 4.9.** For the homothetic motion $\tilde{Q}(u,v) = u \sinh \theta Q_1(u,v)$, the time-like constant slope surface $S$ can be reparametrized by $x(u,v) = \tilde{Q}(u,v) \times h(v)$. Therefore, we have

$$x(u,v) = u \sinh \theta R_Q h(v). \quad (4.8)$$

We can give the following example:

**Example 4.10.** Let us consider a unit space-like curve on $\mathbb{S}_1^2$ defined by

$$h(v) = (0, \cos v, \sin v).$$

Taking $\theta = 7$ and $\coth 7 \cong 1$, the homothetic motion is equal to

$$\tilde{Q}(u,v) = u\sinh 7\left(\cosh(\ln u) + (0, \sinh(\ln u)\sin v, -\sinh(\ln u)\cos v)\right).$$

By using (4.8), we have the following time-like constant slope surface lying in the space-like cone:

$$x(u,v) = u\sinh 7 \begin{bmatrix} \cosh(\ln u) & -\sinh(\ln u)\cos v & -\sinh(\ln u)\sin v \\ -\sinh(\ln u)\cos v & \cosh^2\left(\frac{\ln u}{2}\right) + \sinh^2\left(\frac{\ln u}{2}\right)\cos 2v & \sinh^2\left(\frac{\ln u}{2}\right)\sin 2v \\ -\sinh(\ln u)\sin v & \sinh^2\left(\frac{\ln u}{2}\right)\sin 2v & \cosh^2\left(\frac{\ln u}{2}\right) - \sinh^2\left(\frac{\ln u}{2}\right)\cos 2v \end{bmatrix} \begin{bmatrix} 0 \\ \cos v \\ \sin v \end{bmatrix}$$

and then

$$x(u,v) = \begin{bmatrix} -u\sinh 7 \sinh(\ln u) \\ u\sinh 7 \cosh(\ln u)\cos v \\ u\sinh 7 \cosh(\ln u)\sin v \end{bmatrix}.$$

Thus, we can draw the picture of $x(u,v) = u\sinh\theta R_Q h(v)$ as follows:

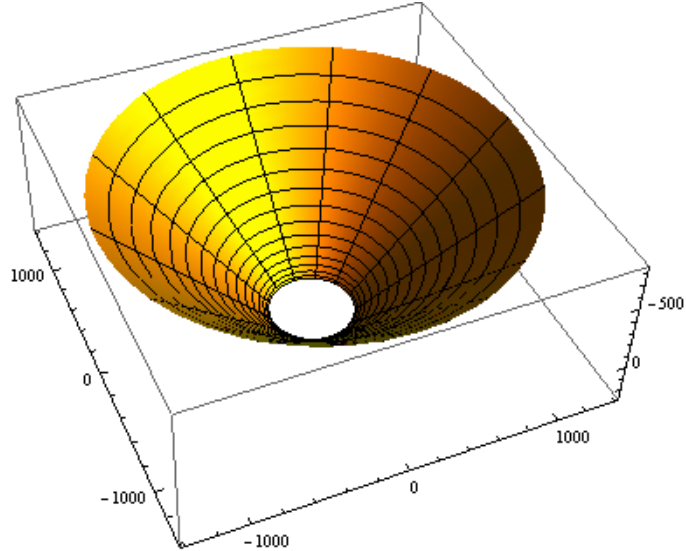

Figure 3: Time-like constant slope surface lying in the space-like cone $x(u,v) = u\sinh\theta R_Q h(v)$, $h(v) = (0, \cos v, \sin v)$, $\theta = 7$.

We can give the following remarks regarding Theorem 4.7 and Corollary 4.9.

**Remark 4.11.** Theorem 4.7 says that the position vectors on the surface $Q_2(u,v)$ are rotated by $Q_1(u,v)$ through the hyperbolic angle $\xi_3(u)$ about the space-like axis span$\{h'(v)\}$.

**Remark 4.10**. Corollary 4.9 says that the position vector of the space-like curve $h(v)$ is rotated by $\tilde{Q}(u,v)$ through the hyperbolic angle $\xi_3(u)$ about the space-like axis span$\{h'(v)\}$ and extended through the homothetic scale $u\sinh\theta$.